\documentclass[pra,aps,amsmath,amssymb,showkeys,twocolumn]{revtex4}
\usepackage{graphicx}
\newcommand{\hh}{{\mathcal{H}}}
\newcommand{\hk}{{\mathcal{K}}}
\newcommand{\pen}{\openone}
\newcommand{\bro}{\boldsymbol{\rho}}

\newcommand{\tr}{{\mathrm{tr}}}

\newcommand{\xdif}{{\mathrm{d}}}
\newcommand{\tpq}{\mathtt{p}}
\newcommand{\tqp}{\mathtt{q}}
\newcommand{\iu}{{\mathtt{i}}}
\newcommand{\veps}{\varepsilon}
\newcommand{\ax}{{\mathsf{X}}}
\newcommand{\hm}{{\mathsf{E}}}
\newcommand{\elm}{{\mathsf{L}}}
\newcommand{\nm}{{\mathsf{N}}}
\newcommand{\cle}{{\mathcal{E}}}
\newcommand{\clt}{{\mathcal{T}}}
\newcommand{\wle}{\widetilde{\mathcal{E}}}
\newcommand{\wte}{\widetilde{\mathcal{T}}}

\newcommand{\weta}{\widetilde{\theta}}
\newcommand{\wel}{\widetilde{\ell}\,}
\newcommand{\niz}{{\mathbf{0}}}
\newcommand{\wbro}{\widetilde{\boldsymbol{\rho}}}

\newcommand{\vcr}{\mathbf{r}}

\begin{document}
\clearpage
\preprint{}

\title{On entropic uncertainty relations for measurements of energy and its ``complement''}

\author{Alexey E. Rastegin}
\affiliation{Department of Theoretical Physics, Irkutsk State University,
Gagarin Bv. 20, Irkutsk 664003, Russia}

\begin{abstract}
Heisenberg's uncertainty principle in application to energy and time
is a powerful heuristics. This statement plays the
important role in foundations of quantum theory and statistical
physics. If some state exists for a finite interval of time, then it
cannot have a completely definite value of energy. It is well known
that the case of energy and time principally differs from more
familiar examples of two non-commuting observables. Since quantum
theory was originating, many approaches to energy-time uncertainties
have been proposed. Entropic way to formulate the uncertainty
principle is currently the subject of active researches. Using the
Pegg concept of complementarity of the Hamiltonian, we obtain
uncertainty relations of the ``energy-time'' type in terms of the
R\'{e}nyi and Tsallis entropies. Although this concept is somehow
restricted in scope, derived relations can be applied to systems
typically used in quantum information processing. Both the
state-dependent and state-independent formulations are of interest.
Some of the derived state-independent bounds are similar to the
results obtained within a more general approach on the basis of
sandwiched relative entropies. The developed method allows us to
address the case of detection inefficiencies.
\end{abstract}


\maketitle

\pagenumbering{arabic}
\setcounter{page}{1}

\section{Introduction}\label{sec1}

The uncertainty principle is widely known among achievements
inspired by quantum physics. The Heisenberg thought experiment
with microscope was first analyzed in qualitative sense
\cite{heisenberg}. As a formal statement, it was explicitly derived
by Kennard \cite{kennard}. This formulation can be extended to
arbitrary pairs of observables \cite{robert}. Robertson's
formulation has later been criticized for several reasons
\cite{deutsch,maass}. There is no general consensus concerning a
proper formulation of the uncertainty principle \cite{lahti}.
Entropic functions provide a powerful tool to characterize
uncertainties in quantum measurements
\cite{ww10,brud11,cbtw17,cerf18}. Other approaches to express
uncertainties in quantum measurements are currently the subject of
interest. In particular, modern investigations concern fine-grained
uncertainty relations \cite{oppwn10,renf13,rafine15,rasene16}, the
sum of variances \cite{huang12,mpati14}, majorization relations
\cite{prz13,fgg13,rpz14,arkz16}, and effective anticommutators
\cite{ktw14}. An important question concerns the role of order in
which measurements have been performed. Traditional formulations deal
with preparation uncertainty relations \cite{rozp17}, since
repeated trials with the same quantum state are assumed. In contrast
to the this scenario, entropic uncertainty relations for successive
measurements were examined \cite{bfs2014,zzhang14,rastadp}. Another
direction focuses on examination of entropic uncertainty relations
from the dynamical viewpoint, including measurements on a system
coupled with bosonic reservoirs \cite{wangh2017,chen2018} and
effects of phase or amplitude damping \cite{chensun18}. The authors
of \cite{wangs2018} studied entropic uncertainty relations for a
particle under the background of a Schwarzschild black hole and its
control.

Considering Heisenberg's thought experiment with microscope, we
conclude that definite momentum cannot be localized in space. In a
similar manner, completely definite value of energy cannot be
localized in time. On the other hand, the case of energy and time
cannot always be treated like the case of usual observables. The
role of time in quantum theory has many facets
\cite{busch90a,busch90b,pegg91,butter2014}. Moreover, there is a
principal reason against the existence of universal form of a
self-adjoint time operator conjugate to the Hamiltonian. In many
basic models, the Hamiltonian spectrum is discrete and the
eigenvalues are bounded from below. This fact is very important from
the physical viewpoint. It is one of well known Pauli's remarks that
a time operator would imply that the Hamiltonian has the entire real
line as its spectrum (see, e.g., footnote 2 in Sec. 8 of English
translation \cite{wpauli80}). Various approaches to energy-time
uncertainty relations are reviewed in \cite{dods15}. First formal
derivations were explicitly given by Mandelstam and Tamm
\cite{mandelstam45} and by Fock and Krylov \cite{fock47}. Further
development of this direction was given in \cite{grab84,miy16}. Such
results were also interpreted as setting a fundamental bound on how
fast any quantum system can evolve \cite{deffner17}. Entropic
uncertainty relations for energy and time can be approached by
constructing an almost-periodic time observable \cite{hall2008}. The
latter is rather inadequate clock for aperiodic systems. The
quantum-clock view on time uncertainty was recently developed in
\cite{colesetal18}.

Despite of many previous attempts, the authors of \cite{nieto1968}
claimed that the proper interpretation of energy-time uncertainty
relations remains to be given. The author of \cite{pegg98}
introduced the concept of complement of the Hamiltonian. This notion
is treated as some quantity that is complementary to the
Hamiltonian. The question has been resolved for a system with
discrete energy levels, for which the ratios of the energy
differences are rational exactly or approximately \cite{pegg98}. It
can be represented by a non-orthogonal resolution of the identity as
well as by an Hermitian operator acting in a suitably extended
space. This approach leads to the commutation relation that is
formally equivalent to the phase-number commutator within the
Pegg--Barnett formalism \cite{PBepl,BP89,PB89}. Using the introduced
quantities, Pegg \cite{pegg98} derived uncertainty relations of the
Robertson type. Although the scope of Pegg's approach is somehow
restricted, it is suitable for many interesting models. Moreover, it
approaches the problem along a direction that is typically used to
motivate impossibility of a Hermitian time operator. Thus, the
notion of energy complement provides an alternate way to understand
``energy-time'' issue. It seems that this approach to energy-time
uncertainty relations has received less attention than it deserves.

The aim of this work is to formulate entropic uncertainty
relations for energy and its complement taken within the Pegg
approach \cite{pegg98}. Our consideration is rather complementary
to entropic uncertainty relations obtained recently in
\cite{colesetal18}. We present entropic uncertainty relations that are
immediately related to measurement statistics. Both the
state-dependent and state-independent formulations will be
addressed. In addition, the developed approach allows one to take into
account the case of detection inefficiencies. On the other hand,
the scope of our results is more restricted as related to the
special case of systems with discrete energy levels. This paper is
organized as follows. The preliminary material is reviewed in
Section \ref{sec2}. Here, we recall the definition of used
entropic functions and describe some details of the Pegg approach
to the problem of energy-time uncertainties. Main results are
presented in Section \ref{sec3}. We derive uncertainty relations
of the Maassen--Uffink type as well as relation with the same
parameter in the corresponding entropies. In Section \ref{sec4},
we conclude the paper with a summary of results obtained.

\section{Preliminaries}\label{sec2}

In this section, we review the required material concerning
generalized entropies. The Pegg concept of complement of the
Hamiltonian will be recalled as well. Let $\tpq=\{p_{i}\}$ be a
discrete probability distribution. For $0<\alpha\neq1$, the
R\'{e}nyi $\alpha$-entropy is defined as \cite{renyi61}
\begin{equation}
R_{\alpha}(\tpq):=\frac{1}{1-\alpha}\>
\ln\!\left(\sum\nolimits_{i}p_{i}^{\alpha}\right)
 . \label{repdf}
\end{equation}
It is known that this entropy does not increase with growth of
$\alpha$. The R\'{e}nyi $\alpha$-entropy is certainly concave for
$\alpha\in(0;1)$. For $\alpha>1$, it is neither purely convex nor
purely concave \cite{ja04}. Here, the situation actually depends
on the dimensionality of probabilistic vectors. For a discussion
of basic properties of (\ref{repdf}), see section 2.7 of
\cite{bengtsson}. In the limit $\alpha\to1$, we have the Shannon
entropy
\begin{equation}
H_{1}(\tpq):=-\sum\nolimits_{i}p_{i}\,\ln{p}_{i}
\, . \label{shadf}
\end{equation}
The limit $\alpha\to\infty$ leads to the so-called min-entropy
\begin{equation}
R_{\infty}(\tpq)=-\ln(\max{p}_{i})
 . \label{minen}
\end{equation}
Tsallis entropies form another important family of generalized
entropies. For $0<\alpha\neq1$, the Tsallis $\alpha$-entropy is
defined as \cite{tsallis}
\begin{equation}
H_{\alpha}(\tpq):=\frac{1}{1-\alpha}\left(\sum\nolimits_{i} p_{i}^{\alpha}-1\right)
=-\sum\nolimits_{i} p_{i}^{\alpha}\,\ln_{\alpha}(p_{i})
\, . \label{tsadf}
\end{equation}
Here, the $\alpha$-logarithm of positive $x$ is given as
$\ln_{\alpha}(x)=\bigl(x^{1-\alpha}-1\bigr)/(1-\alpha)$.
Substituting $\alpha=1$, the right-hand side of (\ref{tsadf})
gives (\ref{shadf}). It will be convenient to introduce norm-like
functionals of discrete probabilistic vectors. For $\beta>0$, we
define
\begin{equation}
\|\tpq\|_{\beta}:=\left(\sum\nolimits_{i}p_{i}^{\beta}\right)^{1/\beta}
\, . \label{norb}
\end{equation}
The right-hand side of (\ref{norb}) gives a legitimate norm only
for $\beta\geq1$. Hence, we can write
\begin{equation}
R_{\alpha}(\tpq):=\frac{\alpha}{1-\alpha}\>
\ln\|\tpq\|_{\alpha}
\, , \qquad
H_{\alpha}(\tpq):=\frac{\|\tpq\|_{\alpha}^{\alpha}-1}{1-\alpha}
\ . \label{retsab}
\end{equation}
For $\alpha>1>\beta>0$, we obviously have
\begin{equation}
\|\tpq\|_{\alpha}\leq1
\, , \qquad
\|\tpq\|_{\beta}\geq1
\, . \label{legr}
\end{equation}

We will also concern differential entropies assigned to a
continuously changed variable. In principle, the formula
(\ref{repdf}) can be rewritten immediately. When the variable of
interest is distributed according to the probability density
function $w(\tau)$, then
\begin{equation}
R_{\alpha}(w):=\frac{1}{1-\alpha}{\ }
\ln\!\left(\int w(\tau)^{\alpha}\,\xdif\tau\right)
 , \label{recon0}
\end{equation}
where $0<\alpha\neq1$. The integral is assumed to be taken over the
interval of values, for which $w(\tau)$ is defined. The
corresponding interval will follow from the context. In the limit
$\alpha\to1$, the expression leads to the differential Shannon
entropy
\begin{equation}
H_{1}(w):=-\int w(\tau)\,\ln{w}(\tau)\,\xdif\tau
\, . \label{difsh}
\end{equation}
It is convenient to extend the notion (\ref{norb}) to the case of
probability density functions. For the given density function
$w(\tau)$ and $\beta>0$, we write
\begin{equation}
\|w\|_{\beta}:=
\left(
\int w(\tau)^{\beta}\,\xdif\tau
\right)^{1/\beta}
 . \label{betynrm}
\end{equation}
In the case of discrete distributions, we deal with (\ref{legr}). It is
provided by the normalization $\|\tpq\|_{1}=1$. On the other hand,
for probability density functions the normalization $\|w\|_{1}=1$
does not provide restrictions analogous to (\ref{legr}). One of
corollaries of this fact is that differential entropies are not
positive definite in general. Formulating the Tsallis version of
uncertainty relations with continuous time, we will use entropies
taken with binning only.

There are several possible ways to fit a quantum counterpart of
generalized entropic functions. The main point is that we deal
here with the case of non-commuting variables. One of existing
approaches is based on the concept of the so-called ``sandwiched''
divergences. In general, the concept of relative entropy, or
divergence, plays the key role in quantum information theory
\cite{preskill,wilde17}. Quantum  relative entropies of the
R\'{e}nyi type are considered as a generalization of this concept.
To resolve the non-commutative case, sandwiched entropies have
found to be useful \cite{mdsft13}. Another approach to
parameterized quantum entropies was thoroughly examined in
\cite{bzhpl16}. The sandwiched R\'{e}nyi relative entropies allow
one to define the corresponding conditional entropies. Using such
entropies, the authors of \cite{colesetal18} formulated entropic
energy-time uncertainty relations with a quantum memory. In the
following, we consider an alternative approach based on the
concept of complement of the Hamiltonian.

Let us proceed to the problem of energy-time uncertainty
relations. Following Einstein, the authors of \cite{MTW73}
emphasized that ``nature provides its own way to localize a point
in spacetime''. That is, coordinates are only convenient but not
preexisting tools. Concrete values of coordinates have no
significance unless the used reference frame is somehow anchored
to certain events. Without a further clarification, our everyday
understanding of the word ``time'' cannot be applied in quantum
scales. Of course, this question is typically asked within the
context of quantum gravity \cite{wigner57,wald89}. On the other hand,
limitations on the accuracy of a quantum clock are closely related
to Heisenberg's uncertainty principle \cite{wigner57}. To simplify
formulas, we will further deal with the units in which $\hbar=1$.
Then the energy scale is inverse to the time scale. The problem of
existence of a proper time operator has found a certain attention
(see, e.g. section III.8 of \cite{holevo82}). For a free
non-relativistic particle with the standard Hamiltonian of kinetic
energy, time representation is built by means of the Fourier
transform. The corresponding entropic uncertainty relation
\cite{grab87} merely repeats the relation of Beckner \cite{beck}
and Bia{\l}ynicki-Birula and Mycielski \cite{birula1}. It is an
improvement of the result derived by Hirschman \cite{hirs}. For
discrete semi-bounded Hamiltonians, the problem was formally
analyzed in \cite{galapon}. The result of \cite{galapon} has been
criticized in \cite{hall2008}. We will use the approach of
Pegg \cite{pegg98} who proposed explicit constructions for
discrete systems with levels of a certain structure.

Let us consider the system with $d+1$ energy levels $\veps_{n}$. It
is convenient to choose the lowest level $\veps_{0}=0$
\cite{pegg98}. We will also assume that energy values are
non-degenerate and numbered in increasing order. The Hamiltonian is
accordingly represented as
\begin{equation}
\hm=\sum\nolimits_{n=0}^{d} \veps_{n}\,|\veps_{n}\rangle\langle\veps_{n}|
\, , \label{hamn}
\end{equation}
where $|\veps_{n}\rangle$ denotes $n$-th energy eigenstate. In the
case of unitary evolution, a pure state changes in time
according to
\begin{align}
&\exp(-\iu\hm\,t)\,|\psi\rangle=
\sum\nolimits_{n=0}^{d} \exp(-\iu\veps_{n}t)\,c_{n}\,|\veps_{n}\rangle
\, , \label{psicn1}\\
&c_{n}=\langle\veps_{n}|\psi\rangle
\, . \nonumber
\end{align}
The author of \cite{pegg98} asked a quantity conjugate to the
Hamiltonian in the sense that $\hm$ is the generator of shifts.
So, one seeks states of the form $|\tau\rangle$, for which
\begin{equation}
\exp(-\iu\hm\,\varDelta\tau)\,|\tau\rangle=|\tau+\varDelta\tau\rangle
\, . \label{taush}
\end{equation}
It is not difficult to get the final expression \cite{pegg98}
\begin{equation}
|\tau\rangle=\frac{1}{\sqrt{d+1}}
\sum_{n=0}^{d} \exp(-\iu\veps_{n}\tau)\,|\veps_{n}\rangle
\, . \label{taush1}
\end{equation}
Such expressions are typical in considering eigenstates of
complementary observables in finite dimensions. For equidistant
levels, we will deal just with two complementary observables. In
the context of uncertainty relations, this question was analyzed
in \cite{kraus87}. The main question is how to treat the case of
unequally spaced energy levels \cite{pegg98}.

The parameter $\tau$ in (\ref{taush1}) can be varied continuously.
Thus, we have arrived at an over-complete set of kets of the form
(\ref{taush1}). It is generally impossible to build an orthonormal
basis of such states \cite{pegg98}. Nevertheless, one is able to get
a non-orthogonal resolution of the identity on $\hh_{d+1}$. Suppose
that the ratios $\veps_{n}/\veps_{1}$ are rational numbers or can be
sufficiently closely approximated by them. For the former, we write
\begin{equation}
\frac{\veps_{n}}{\veps_{1}}=\frac{B_{n}}{A_{n}}
\ , \label{rarat}
\end{equation}
where integers $B_{n}$ and $A_{n}$ are mutually prime. By $r_{1}$,
one denotes the lowest common multiple of the values of $A_{n}$
for $n>1$. Defining $r_{0}=0$ and $r_{n}=r_{1}B_{n}/A_{n}$ for
$n>1$, we deal with integer numbers $r_{n}$. This results in the
formula
\begin{equation}
\veps_{n}=\frac{2\pi{r}_{n}}{T_{c}}
\ , \label{perte}
\end{equation}
where $T_{c}=2\pi{r}_{1}/\veps_{1}$. Following \cite{pegg98}, we
take $s+1$ states of the form (\ref{taush1}) for the values
\begin{equation}
\tau_{m}=\tau_{0}+m\,\frac{T_{c}}{s+1}
\qquad (m=0,1,\ldots,s)
\, . \label{taum}
\end{equation}
The intermediate values $\tau_{1},\ldots,\tau_{s}$ are uniformly
distributed between the points $\tau_{0}$ and $\tau_{0}+T_{c}$. As
was shown in \cite{pegg98}, one finally gets
\begin{equation}
\frac{d+1}{s+1}\sum_{m=0}^{s}|\tau_{m}\rangle\langle\tau_{m}|
=\pen_{d+1}
\, , \label{ropmt}
\end{equation}
where $\pen_{d+1}$ is the identity operator on $\hh_{d+1}$.
Therefore, we have arrived at a non-orthogonal resolution of the
identity for measuring an energy complement. The relation
(\ref{ropmt}) is satisfied exactly when the ratios
$\veps_{n}/\veps_{1}$ are rational and the differences
$r_{\ell}-r_{n}$ are not multiplies of $s+1$. One can ensure the
latter by choosing $s+1>\max{r}_{n}$. In other respects, we have a
freedom in the choice of $s\geq{d}$. If these ratios are
irrational but sufficiently well approximated by rational numbers,
the relation (\ref{ropmt}) holds up to a negligible additive term
\cite{pegg98}. To each energy level $\veps_{n}$, we can assign the
natural period $2\pi/\veps_{n}$. When $\veps_{n}/\veps_{1}$ are
exact rational numbers, the characteristic time $T_{c}$ has a
simple physical interpretation. It represents the smallest
non-zero time taken for the system to return to its initial state
\cite{pegg98}. Hence, the state $|\tau+T_{c}\rangle$ will coincide
with $|\tau\rangle$. Focusing on the corresponding range in
(\ref{taum}) prevents us from including the same state twice or
more.

Taking a positive operator-valued measure (POVM), we still
not reach an observable represented by a Hermitian operator. On
the other hand, uncertainties themselves are rather connected with
a spread of probability distribution. In this regard, the entropic
way to formulate uncertainty relations is quite sufficient since
entropies are immediately calculated for concrete values of
probabilities. Although the question of building the complement
observable can be resolved within Naimark's extension
\cite{pegg98}, we can express entropic uncertainty relations
without it.

Measuring the energy, we use projection-valued measure
$\cle=\bigl\{|\veps_{n}\rangle\langle\veps_{n}|\bigr\}$. To the
given state $\bro$, we assign the probabilities
$\langle\veps_{n}|\bro|\veps_{n}\rangle$. By
$R_{\alpha}(\cle;\bro)$ and $H_{\alpha}(\cle;\bro)$, we denote the
$\alpha$-entropies (\ref{repdf}) and (\ref{tsadf}) calculated with
these probabilities. The complement of energy is described by
rank-one POVM
$\clt=\bigl\{|\theta_{m}\rangle\langle\theta_{m}|\bigr\}$, where
\begin{equation}
|\theta_{m}\rangle=\sqrt{\frac{d+1}{s+1}}\>|\tau_{m}\rangle
\, . \label{thetm}
\end{equation}
To the prepared state $\bro$, we assign the entropies
$R_{\alpha}(\clt;\bro)$ and $H_{\alpha}(\clt;\bro)$ calculated
according to the probabilities
$\langle\theta_{m}|\bro|\theta_{m}\rangle$. In many respects, the
above construction is similar to the Pegg--Barnett formalism
\cite{PBepl,BP89,PB89}. This formalism allows us to fit a Hermitian
operator to represent quantum phase. Since Dirac's famous work
\cite{dirac27} on quantum electrodynamics had appeared, the quantum
phase problem has been studied from different viewpoints
\cite{nieto1968,lynch}. An intuitive assumption is that the
operators of optical phase and photon number are canonically
conjugate. Instead of using the infinite Hilbert space from the
begin, the Pegg--Barnett formalism deals with a finite but
arbitrarily large state space \cite{PBepl,BP89}. The final step is
to find the limit of desired quantities as the dimensionality tends
to infinity. The authors of \cite{pvb90} have developed this
approach with respect to the concept of canonical conjugacy.

\section{Main results}\label{sec3}

In this section, we derive various forms of entropic uncertainty
relations for energy and its complement. Let us begin with
entropic uncertainty relations of the Maassen--Uffink type.
Following \cite{ramubs13}, we introduce the quantity
\begin{equation}
g(\cle,\clt;\bro):=\max
\frac{\bigl|\langle\veps_{n}|\theta_{m}\rangle\,\langle\theta_{m}|\bro|\veps_{n}\rangle\bigr|}
{\langle\veps_{n}|\bro|\veps_{n}\rangle^{1/2}\,\langle\theta_{m}|\bro|\theta_{m}\rangle^{1/2}}
\ , \label{getdf}
\end{equation}
where the maximization is performed under the conditions
$\langle\veps_{n}|\bro|\veps_{n}\rangle\neq0$ and
$\langle\theta_{m}|\bro|\theta_{m}\rangle\neq0$. In general, the
quantity (\ref{getdf}) depends on the used construction of states
$|\theta_{m}\rangle$. To get uncertainty relations in terms of
R\'{e}nyi entropies, purely algebraic operations are required. The
case of Tsallis entropies is not so immediate. We will use the
method of \cite{rast11u}, where the minimization problem was
examined. Entropic uncertainty relations for energy and its
complement are posed as follows. For any prepared state $\bro$, we
have
\begin{align}
R_{\alpha}(\cle;\bro)+R_{\beta}(\clt;\bro)&\geq-2\ln{g}(\cle,\clt;\bro)
\, , \label{rengr}\\
H_{\alpha}(\cle;\bro)+H_{\beta}(\clt;\bro)&\geq\ln_{\mu}\!\left\{g(\cle,\clt;\bro)^{-2}\right\}
\, , \label{tsagr}
\end{align}
where positive entropic parameters obey $1/\alpha+1/\beta=2$ and
$\mu=\max\{\alpha,\beta\}$. The condition $1/\alpha+1/\beta=2$
reflects the fact that the Maassen--Uffink result is based on
Riesz's theorem \cite{riesz27}. An alternative viewpoint is that
the above uncertainty relations follow from the monotonicity of
the quantum relative entropy \cite{ccyz12}. For an arbitrary
choice of $\alpha$ and $\beta$, the problem of obtaining general
entropic bounds was examined in \cite{zbp2014}.

The inequalities (\ref{rengr}) and (\ref{tsagr}) are preparation
uncertainty relations formulated in terms of both the R\'{e}nyi
and Tsallis entropies. Due to (\ref{getdf}), these entropic bounds
depend on the way in which we have built the POVM
$\clt=\bigl\{|\theta_{m}\rangle\langle\theta_{m}|\bigr\}$. This
POVM is constructed of kets that are mutually unbiased with the
eigenstates of the Hamiltonian. A certain freedom takes place in
the choice of actual referent values $\tau_{m}$ of time. Thus, we
obtained a kind on entropic ``energy-time'' uncertainty relations.
It is not insignificant that our relations directly connect to
measurement statistics. In this regard, they differ from entropic
uncertainty relations derived in \cite{colesetal18}. Another
distinction is that both the bounds (\ref{rengr}) and
(\ref{tsagr}) explicitly depend on the measured state $\bro$.

As was explained in \cite{rast11u}, the state-dependent
uncertainty bounds can be converted into a state-independent form.
It turned out that state-independent entropic bounds are expressed
in terms of $s$ solely. To do so, we merely write
\begin{equation}
g(\cle,\clt;\bro)\leq{f}(\cle,\clt):=\max\,\bigl|\langle\veps_{n}|\theta_{m}\rangle\bigr|
\, . \label{fetdf}
\end{equation}
This inequality follows from combining (\ref{getdf}) with the
Cauchy--Schwarz inequality. It is easy to check that
$f(\cle,\clt)=(s+1)^{-1/2}$ according to the chosen $s$. As a
result, we obtain
\begin{align}
R_{\alpha}(\cle;\bro)+R_{\beta}(\clt;\bro)&\geq\ln(s+1)
\, , \label{renfr}\\
H_{\alpha}(\cle;\bro)+H_{\beta}(\clt;\bro)&\geq\ln_{\mu}(s+1)
\, , \label{tsafr}
\end{align}
where $1/\alpha+1/\beta=2$ and $\mu=\max\{\alpha,\beta\}$. In
other words, the entropic bounds (\ref{renfr}) and (\ref{tsafr})
are expressed in terms of the number $s+1$ of the reference
instants of time. Here, we see a similarity to the entropic bound
in energy-time uncertainty relations given in \cite{colesetal18}.
The following fact should be pointed out. Since the states
$|\theta_{m}\rangle$ are sub-normalized, the corresponding
probabilities cannot reach $1$. Hence, the entropies
$R_{\beta}(\clt;\bro)$ and $H_{\beta}(\clt;\bro)$ in the above
relations are certainly non-zero. Using
$\langle\theta_{m}|\bro|\theta_{m}\rangle\leq(d+1)/(s+1)$, we
easily obtain
\begin{equation}
R_{\beta}(\clt;\bro)\geq\ln\!\left(\frac{s+1}{d+1}\right)=:\varGamma
\, . \label{vgadf}
\end{equation}
Of course, this estimation from below is only approximate.
Nevertheless, it can be used to characterize an unavoidable amount
of uncertainty in the measurement $\clt$. Subtracting
$\varGamma$ from both the sides of (\ref{renfr}), we have
\begin{equation}
R_{\alpha}(\cle;\bro)+R_{\beta}(\clt;\bro)-\varGamma\geq\ln(d+1)
\, . \label{srenfr}
\end{equation}
After subtracting, the entropic lower bound is determined by the
logarithm of dimensionality for every choice of time moments.

Using the entropic approach, we can take into account possible
inefficiencies of the detectors used. Since measurement devices
inevitably suffer from losses, the ``no-click'' probability is
non-zero in practice. Here, we consider the following model. Let
the parameter $\eta\in[0;1]$ characterize a detector efficiency.
To the given value $\eta$ and probability distribution
$\tpq=\{p_{i}\}$, we assign a ``distorted'' distribution
$\tpq^{(\eta)}$ such that
\begin{equation}
p_{i}^{(\eta)}=\eta{p}_{i}
\ , \qquad
p_{\varnothing}^{(\eta)}=1-\eta
\ . \label{petad}
\end{equation}
The probability $p_{\varnothing}^{(\eta)}$ corresponds to the
no-click event. The above formulation is inspired by the first
model of detection inefficiencies used by the authors of
\cite{rchtf12} for cycle scenarios of the Bell type. For the sake
of simplicity, we restrict a consideration to the Shannon
entropies. It was mentioned in \cite{ramubs13} that
\begin{equation}
H_{\alpha}\bigl(\tpq^{(\eta)}\bigr)=\eta^{\alpha}H_{\alpha}(\tpq)+h_{\alpha}(\eta)
\, , \label{qtlm0}
\end{equation}
where the binary Tsallis entropy $h_{\alpha}(\eta)$ reads as
\begin{equation}
h_{\alpha}(\eta)=-\,\eta^{\alpha}\ln_{\alpha}(\eta)-(1-\eta)^{\alpha}\ln_{\alpha}(1-\eta)
\, . \label{bnta}
\end{equation}
For the Shannon entropies, one gives
\begin{equation}
H_{1}\bigl(\tpq^{(\eta)}\bigr)=\eta\,H_{1}(\tpq)+h_{1}(\eta)
\, . \label{adbin}
\end{equation}
In the case considered, we have
\begin{align}
H_{1}(\cle^{(\eta_{\cle})};\bro)+H_{1}(\clt^{(\eta_{\clt})};\bro)
&\geq
-2\eta\ln{g}(\cle,\clt;\bro)+2h_{1}(\eta)
\nonumber\\
&\geq\eta\ln(s+1)+2h_{1}(\eta)
\, . \label{etaun}
\end{align}
By $\eta=\min\{\eta_{\cle},\eta_{\clt}\}\geq1/2$, we mean the
minimum of the two efficiencies corresponding respectively to
measurements of energy and its complement. We see that detector
inefficiencies will produce additional uncertainties in the
entropies of actually measured data. For low values of the
efficiency, measurement statistics will mainly reflect
detector-generated uncertainties.

Since the states $|\theta_{m}\rangle$ lead to a non-orthogonal
resolution of the identity in $\hh_{d+1}$, they cannot be
eigenstates of a Hermitian operator acting in this space. On the
other hand, any POVM-measurement can be realized as a projective
one in suitably extended space. In principle, this possibility is
established by the Naimark theorem. Its general discussion with
applications can be found in \cite{holevo82,helstrom76}. It is
sufficient for our aims to focus on the case of rank-one POVMs.
Then the corresponding projective measurement may be constructed
in a simplified manner as follows (see, e.g., section 3.1 of
\cite{preskill}). Components of $s+1$ kets $|\theta_{m}\rangle$
are treated as elements of certain $(d+1)\times(s+1)$ matrix.
Adding this matrix by suitable number of rows, we can obtain a
unitary matrix of size $s+1$. Each Hermitian operator acting in
the extended space $\hh_{s+1}=\hh_{d+1}\oplus\hk$ will have
$s+1$ eigenstates. The energy basis will include extra states, so
that we obtain an orthogonal resolution
$\wle_{s+1}=\bigl\{|\wel\rangle\langle\wel|\bigr\}_{\ell=0}^{s}$.
Following \cite{pegg98}, we consider kets of the form
\begin{equation}
|\weta_{m}\rangle=\frac{1}{\sqrt{s+1}}
\sum_{\ell=0}^{s} \exp(-\iu\ell\,\theta_{m})\,|\wel\rangle
\, . \label{weta1}
\end{equation}
In this way, we obtain another orthogonal resolution
$\wte_{s+1}=\bigl\{|\weta_{m}\rangle\langle\weta_{m}|\bigr\}_{m=0}^{s}$.
It must be stressed that the original energy eigenstates
$|\veps_{n}\rangle$ with $n=0,1,\ldots,d$ are rearranged so that
\begin{equation}
|\wel\rangle=|\veps_{n}\rangle\oplus\niz
\, , \label{weln}
\end{equation}
whenever $\ell=r_{n}$. The latter is possible due to
$s+1>\max{r}_{n}$. If $\ell\neq{r}_{n}$ for all $n=0,1,\ldots,d$,
then the ket $|\wel\rangle$ has non-zero components only in $\hk$.
In (\ref{weta1}), the numbers $\theta_{m}$ are defined as
\cite{pegg98}
\begin{equation}
\theta_{m}=\frac{2\pi\tau_{m}}{T_{c}}
\ . \label{thtmd}
\end{equation}
These numbers lie in a range of length $2\pi$ between $\theta_{0}$
and $\theta_{0}+2\pi$, where $\theta_{0}=2\pi\tau_{0}/T_{c}$. To
each density matrix $\bro$ on $\hh_{d+1}$, we assign the matrix
$\wbro$ of size $s+1$ by adding zero rows and columns. Obviously, we
have
\begin{equation}
\langle\weta_{m}|\wbro|\weta_{m}\rangle=\langle\theta_{m}|\bro|\theta_{m}\rangle
\label{thmth}
\end{equation}
for all $m$,
$\langle\wel|\wbro|\wel\rangle=\langle\veps_{n}|\bro|\veps_{n}\rangle$
for $\ell=r_{n}$, and $\langle\wel|\wbro|\wel\rangle=0$ for
$\ell\neq{r}_{n}$. In the case considered, we introduce the
following two operators,
\begin{equation}
\sum_{\ell=0}^{s} \ell\,|\wel\rangle\langle\wel|
\, , \qquad
\sum_{m=0}^{s} \theta_{m}\,|\weta_{m}\rangle\langle\weta_{m}|
\, . \label{twoho}
\end{equation}
Up to a factor, the former operator gives a Hamiltonian acting in
the extended space. The second one is formally equivalent to the
operator of optical phase due to Pegg and Barnett. For the above
operators, one can easily obtain uncertainty relations of the
Robertson type. Their discussion together with the limiting case
$s\to\infty$ can be found in \cite{pegg98}. On the other hand,
entropic uncertainty relations are rather connected with
resolutions of the identity. In this sense, we will mainly focus
on probability distributions and, after taking the limit,
probability density functions.

The author of \cite{pegg98} also mentioned how to unify the
approach for all systems of the type considered. We can examine
basic quantities in the limit $s\to\infty$. As the difference
between successive values of $\tau_{m}$ tends to zero, the
probability to lie in the small range between $\tau$ and
$\tau+\varDelta\tau$ is equal to $w_{\bro}(\tau)\,\varDelta\tau$.
Here, we define
\begin{equation}
w_{\bro}(\tau)=\frac{\langle\tau^{\prime}|\bro|\tau^{\prime}\rangle}{T_{c}}
\ , \label{wbta0}
\end{equation}
in terms of rescaled kets
$|\tau^{\prime}\rangle=\sqrt{d+1}\,|\tau\rangle$. Taking
$\varDelta\tau=T_{c}\,/(s+1)$, the function (\ref{wbta0})
satisfies
\begin{equation}
w_{\bro}(\tau_{m})\,\varDelta\tau=\langle\theta_{m}|\bro|\theta_{m}\rangle
\, . \label{wbta1}
\end{equation}
In line with (\ref{thtmd}), we also have the relation
$U_{\wbro}(\theta_{m})\,\varDelta\theta=w_{\bro}(\tau_{m})\,\varDelta\tau$
with $\varDelta\theta=2\pi/(s+1)$. Here, the density matrix
$\wbro$ is assumed to be obtained from $\bro$ by adding zero rows
and columns. According to (\ref{recon0}), we introduce
differential R\'{e}nyi $\alpha$-entropies $R_{\alpha}(w_{\bro})$
and $R_{\alpha}\bigl(U_{\wbro}\bigr)$. In contrast to entropies of
discrete probability distributions, differential entropies may
take negative values. Hence, uncertainty relations in terms of
differential Tsallis entropies cannot be treated similarly to
(\ref{tsafr}). As the method of appendix of \cite{rast11u} uses
(\ref{legr}), we will apply resulting relations to Tsallis
entropies with binning.

Let positive parameters $\alpha$ and $\beta$ satisfy the condition
$1/\alpha+1/\beta=2$. To consider the limit $s\to\infty$, we treat
probability distributions as related to the extended space
$\hh_{s+1}$. As was mentioned above, the observables (\ref{twoho})
are canonically conjugate in the sense of the Pegg--Barnett
formalism. For probabilistic vectors $\tpq=\{p_{n}\}$ with
$p_{n}=\langle\veps_{n}|\bro|\veps_{n}\rangle$ and
$\tqp=\{q_{m}\}$ with
$q_{m}=\langle\theta_{m}|\bro|\theta_{m}\rangle$, we have
\begin{align}
\|\tpq\|_{\alpha}&\leq
\left(\frac{1}{s+1}\right)^{(1-\beta)/\beta}\|\tqp\|_{\beta}
\, , \nonumber\\
\|\tqp\|_{\alpha}&\leq
\left(\frac{1}{s+1}\right)^{(1-\beta)/\beta}\|\tpq\|_{\beta}
\, , \label{twipq}
\end{align}
where $1/2<\beta<1<\alpha$. The formulas (\ref{twipq}) follow from
the Riesz theorem. The limiting procedure results in the
probability density function, so that $q_{m}$ is finally replaced
with $U_{\wbro}(\theta_{m})\,\xdif\theta$. Here, we can write
\begin{align}
\|\tpq\|_{\alpha}&\leq
\left(\frac{1}{2\pi}\right)^{(1-\beta)/\beta}\bigl\|U_{\wbro}\bigr\|_{\beta}
\, , \nonumber\\
\bigl\|U_{\wbro}\bigr\|_{\alpha}&\leq
\left(\frac{1}{2\pi}\right)^{(1-\beta)/\beta}\|\tpq\|_{\beta}
\, , \label{tiup}
\end{align}
where $1/\alpha+1/\beta=2$ and $1/2<\beta<1<\alpha$. These
relations can be derived similarly to the method of the paper
\cite{numph12}. The latter is devoted to number-phase uncertainty
relations in terms of generalized entropies. Differential
entropies are calculated with probability density functions that
depend on rescaling of the random variable. It is better to do
this step in terms of norm-like functionals. Combining
$U_{\wbro}(\theta)\,\xdif\theta=w_{\bro}(\tau)\,\xdif\tau$ with
$\theta=2\pi\tau/T_{c}$, we also obtain
\begin{equation}
\bigl\|U_{\wbro}\bigr\|_{\beta}=\left(\frac{2\pi}{T_{c}}\right)^{(1-\beta)/\beta}\|w_{\bro}\|_{\beta}
\, . \label{ubwb}
\end{equation}
Hence, the ``twin'' relations (\ref{tiup}) are rewritten as
\begin{align}
\|\tpq\|_{\alpha}&\leq
\left(\frac{1}{T_{c}}\right)^{(1-\beta)/\beta}\|w_{\bro}\|_{\beta}
\, , \nonumber\\
\|w_{\bro}\|_{\alpha}&\leq
\left(\frac{1}{T_{c}}\right)^{(1-\beta)/\beta}\|\tpq\|_{\beta}
\, , \label{twip}
\end{align}
under the same conditions on $\alpha$ and $\beta$. Using simple
algebraic operations, we convert (\ref{twip}) into entropic
uncertainty relations with continuous time, viz.
\begin{equation}
R_{\alpha}(\cle;\bro)+R_{\beta}(w_{\bro})
\geq\ln{T}_{c}
\, , \label{ctren}
\end{equation}
where $1/\alpha+1/\beta=2$. The obtained entropic bound is very
similar to the bound given in \cite{colesetal18}. It seems that
entropic bounds of such a kind are different manifestations of the
same fundamental restriction. Note that our relation deals with
entropic functions directly related to measurement statistics. In
this sense, one characterizes energy-time uncertainties in a very
traditional style. Thus, we have obtained an old-fashioned
counterpart of entropic energy-time relations proposed in
\cite{colesetal18}.

Since the right-hand side of (\ref{ctren}) involves a dimensional
parameter, there is a dependence on the chosen unit of time. On
the other hand, differential entropy $R_{\beta}(w_{\bro})$ also
depends on the time unit. The mentioned dependence is such that
rescaling time will contribute the additive term to both the sides
of the relation (\ref{ctren}). In this sense, our entropic
uncertainty relations with continuous time are independent of the
time unit. To get a dimensionless formulation explicitly, we can
consider entropic uncertainty relations with time binning. The
interval $[\tau_{0};\tau_{0}+T_{c}]$ is divided into the set of
bins between some ordered marks $\tau_{j}$. In contrast to the
case (\ref{taum}), these values can generally be chosen in
arbitrary way. By $\delta\tau$, we mean the maximum of the
differences $\tau_{j+1}-\tau_{j}$. Instead of $w_{\bro}(\tau)$, we
now deal with probabilities of the form
\begin{equation}
q_{j}^{(\delta)}:=\int\nolimits_{\tau_{j}}^{\tau_{j+1}}
w_{\bro}(\tau)\,\xdif\tau
\, , \label{qdwt}
\end{equation}
resulting in the discrete distribution $\tqp_{\,\clt}^{(\delta)}$.
Due to (\ref{twip}), we obtain the inequalities
\begin{align}
\|\tpq\|_{\alpha}&\leq
\left(\frac{\delta\tau}{T_{c}}\right)^{(1-\beta)/\beta}\bigl\|\tqp_{\,\clt}^{(\delta)}\bigr\|_{\beta}
\, , \nonumber\\
\bigl\|\tqp_{\,\clt}^{(\delta)}\bigr\|_{\alpha}&\leq
\left(\frac{\delta\tau}{T_{c}}\right)^{(1-\beta)/\beta}\|\tpq\|_{\beta}
\, , \label{dwip}
\end{align}
where $1/\alpha+1/\beta=2$ and $1/2<\beta<1<\alpha$. Details of
deriving (\ref{dwip}) from (\ref{twip}) are quite similar to that
was given in section 3.3 of \cite{fprast15}. Using (\ref{dwip}),
we finally obtain
\begin{align}
R_{\alpha}(\cle;\bro)+R_{\beta}\bigl(\tqp_{\,\clt}^{(\delta)};\bro\bigr)
&\geq\ln\!\left(\frac{T_{c}}{\delta\tau}\right)
 , \label{crbin}\\
H_{\alpha}(\cle;\bro)+H_{\beta}\bigl(\tqp_{\,\clt}^{(\delta)};\bro\bigr)
&\geq\ln_{\mu}\!\left(\frac{T_{c}}{\delta\tau}\right)
 , \label{ctbin}
\end{align}
where $1/\alpha+1/\beta=2$ and $\mu=\max\{\alpha,\beta\}$. The
inequalities (\ref{crbin}) and (\ref{ctbin}) give entropic
uncertainty relations with time binning. As was
already mentioned, the actual bins can be chosen irrespectively to
(\ref{taum}). In this sense, uncertainty relations of
``energy-time'' kind are written in unifying way, when the system
considered is characterized by the single parameter $T_{c}$. Of
course, the above results are derived under assumptions used
initially in building the POVM
$\clt=\bigl\{|\theta_{m}\rangle\langle\theta_{m}|\bigr\}$.

\begin{figure*}
\centering \includegraphics[height=7.4cm]{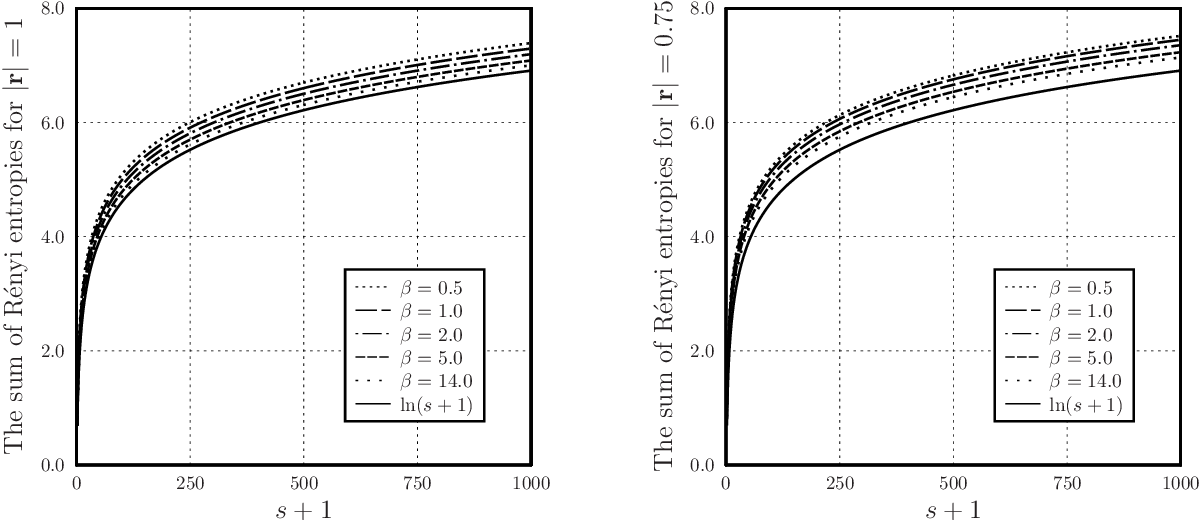}
\caption{\label{fig1} The left-hand side of (\ref{renfr}) for
several values of $\beta$, with $|\vcr|=1$ on the left plot and $|\vcr|=0.75$
on the right one.}
\end{figure*}

Let us consider an example of preparation uncertainty relations for
energy and its complement. The simplest case deals with repeated
measurements on a single qubit. It can be meant as a spin-$1/2$
particle in an external magnetic field. The Hamiltonian is
proportional to the $z$-Pauli matrix. However, we recall that the
energy scale should be shifted to provide $\veps_{0}=0$. The latter
is required to construct POVMs
$\clt=\bigl\{|\theta_{m}\rangle\langle\theta_{m}|\bigr\}$. The
number $s+1$ of referent moments changes from $2$ up to infinity. In
this example, we may simply put $\tau_{0}=0$. It is usual to
represent qubit states by vectors of the Bloch ball. In Fig.
\ref{fig1}, we plot the left-hand side of (\ref{renfr}) together
with lower bound $\ln(s+1)$ for several values of $\beta$. The Bloch
vector $\vcr$ points out along the $x$-axis, whereas its modulus is
taken to be $|\vcr|=1$ and $|\vcr|=0.75$. For equatorial qubit states, the
entropy $R_{\alpha}(\cle;\bro)$ is constant. Thus, the curves mainly
reflect changes in $R_{\beta}(\clt;\bro)$. The abscissa includes
values of $s+1$ between $2$ and $1000$, whence a pass to the case of
continuous time with binning seems to be clear. All the curves lie
near $\ln(s+1)$, so the state-independent lower bound is
sufficiently tight. When $|\vcr|$ decreases, the curves become more
closely to each other, though they slightly shift upward. To take
into account this small increase, we can consider the
state-dependent relation (\ref{rengr}).

Using the treatment of measurements in $\hh_{s+1}$, we can obtain
entropic uncertainty relations of another type. By construction, the
two bases $\bigl\{|\wel\rangle\bigr\}_{\ell=0}^{s}$ and
$\bigl\{|\weta_{m}\rangle\bigr\}_{m=0}^{s}$ are mutually unbiased.
Hence, we can write entropic uncertainty relations for MUBs derived
in \cite{molmer09} and later extended \cite{ramubs13}. If the
density matrix $\wbro$ is obtained from $\bro$ by adding zero
components, then
\begin{equation}
R_{\alpha}(\cle;\bro)=R_{\alpha}\bigl(\wle;\wbro\bigr)
\, , \qquad
R_{\alpha}(\clt;\bro)=R_{\alpha}\bigl(\wte;\wbro\bigr)
\, , \label{entrer}
\end{equation}
and similarly for the Tsallis entropies. By suitable substitutions
into formulas (17) and (18) of \cite{ramubs13}, for
$\alpha\in(0;2]$ one gets
\begin{align}
H_{\alpha}(\cle;\bro)+H_{\alpha}(\clt;\bro)
&\geq2\ln_{\alpha}\!\left(\frac{2s+2}{(s+1)\,\tr(\bro^{2})+1}\right)
\nonumber\\
&\geq2\ln_{\alpha}\!\left(\frac{2s+2}{s+2}\right)
 . \label{2muh}
\end{align}
Uncertainty relations for MUBs in terms of R\'{e}nyi entropies were
presented in \cite{ramubs13} and later improved in \cite{rastosid}.
Applying the results of \cite{ramubs13,rastosid} to the case
considered, for $\alpha\geq2$ we obtain
\begin{align}
&R_{\alpha}(\cle;\bro)+R_{\alpha}(\clt;\bro)\geq
\frac{2}{\alpha-1}\>\ln\!\left(\frac{2s+2}{(s+1)\,\tr(\bro^{2})+1}\right)
\nonumber\\
&+\frac{2\alpha-4}{\alpha-1}\>
\ln\!\left(\frac{\sqrt{2}\,s+\sqrt{2}}{\sqrt{s(s+1)\,\tr(\bro^{2})-s}+\sqrt{2}}\right)
 . \label{2mur}
\end{align}
In particular, the corresponding min-entropies obey
\begin{align}
&R_{\infty}(\cle;\bro)+R_{\infty}(\clt;\bro)
\nonumber\\
&\geq
2\ln\!\left(\frac{\sqrt{2}\,s+\sqrt{2}}{\sqrt{s(s+1)\,\tr(\bro^{2})-s}+\sqrt{2}}\right)
 . \label{2murm}
\end{align}
Thus, we have obtained state-dependent uncertainty relation in
terms of both the Tsallis and R\'{e}nyi entropies. The derived
bounds are expressed in terms of purity $\tr(\bro^{2})$. The above
expressions are especially useful, when purity of the measured
state is sufficiently far from $1$. For the case of pure states,
the results (\ref{2mur}) and (\ref{2murm}) are used with
$\tr(\bro^{2})=1$. For instance, the  min-entropies satisfy
\begin{equation}
R_{\infty}\bigl(\cle;|\psi\rangle\langle\psi|\bigr)+
R_{\infty}\bigl(\clt;|\psi\rangle\langle\psi|\bigr)\geq
2\ln\!\left(\frac{\sqrt{2}\,s+\sqrt{2}}{s+\sqrt{2}}\right)
 . \label{2murmp}
\end{equation}
Of course, the latter remains valid for arbitrary state. Using the
results of \cite{imai07}, we can improve (\ref{2murmp}). By
$\|\ax\|_{\infty}$, we mean the spectral norm of the
operator $\ax$. It is defined as the maximal singular value of
$\ax$. Let $\elm$ and $\nm$ be positive operators that satisfy
$\elm\leq\pen_{d+1}$ and $\nm\leq\pen_{d+1}$; then \cite{imai07}
\begin{equation}
\tr(\elm\bro)+\tr(\nm\bro)\leq1+\bigl\|\sqrt{\elm}\,\sqrt{\nm}\,\bigr\|_{\infty}
\, . \label{immy0}
\end{equation}
This results generalizes an inequality mentioned in \cite{dvsr05}
for measurements in two orthonormal bases. The authors of
\cite{imai07} used (\ref{immy0}) to derive generalized uncertainty
relations of the Landau--Pollak type. Substituting
$\elm=|\theta_{m}\rangle\langle\theta_{m}|$ and
$\nm=|\veps_{n}\rangle\langle\veps_{n}|$ gives
\begin{equation}
\bigl\|\sqrt{\elm}\,\sqrt{\nm}\,\bigr\|_{\infty}=\frac{1}{\sqrt{s+1}}
\, . \label{immy1}
\end{equation}
We now combine (\ref{immy0}) with (\ref{immy1}) and also take into
account $q_{m}(\clt;\bro)\leq(d+1)/(s+1)$. Together, these
observations lead to
\begin{align}
&\max{p}_{n}(\cle;\bro)+\max{q}_{m}(\clt;\bro)
\leq1+\Upsilon
\, , \nonumber\\
&\Upsilon:=\min\left\{\frac{1}{\sqrt{s+1}}\,,\frac{d+1}{s+1}\right\}
 . \label{immy2}
\end{align}
When $s+1>(d+1)^{2}$, we have $\Upsilon=(d+1)/(s+1)$. We
also note that the function $x\mapsto-\,\ln{x}$ is convex and
decreasing. Combining these points with (\ref{minen}) and
(\ref{immy2}), one gets
\begin{equation}
R_{\infty}(\cle;\bro)+R_{\infty}(\clt;\bro)\geq
2\ln\!\left(\frac{2}{1+\Upsilon}\right)
 . \label{murmp}
\end{equation}
For large $s$, the right-hand side of (\ref{murmp}) is
approximately equal to $\ln4$. In the same limit, the right-hand
side of (\ref{2murm}) becomes $\ln2-\ln\bigl(\tr(\bro^{2})\bigr)$.
Applying the latter to the completely mixed state, we obtain the
lower bound $\ln(2d+2)$. When we consider low-purity states of a
system with several energy levels, the formula (\ref{2murm}) is
better than (\ref{murmp}). In other cases, the result
(\ref{murmp}) seems to be preferable.

Using Tsallis entropies with the same parameter $\alpha$, we can
again address the case of detection inefficiencies. It is natural
to suppose that both the efficiencies $\eta_{\cle}$ and
$\eta_{\clt}$ are not less than $1/2$. Due to (\ref{qtlm0}) and
(\ref{2muh}), one gets
\begin{align}
&H_{\alpha}(\cle^{(\eta_{\cle})};\bro)+H_{\alpha}(\clt^{(\eta_{\clt})};\bro)
\nonumber\\
&\geq
2\eta^{\alpha}\ln_{\alpha}\!\left(\frac{2s+2}{(s+1)\,\tr(\bro^{2})+1}\right)
+2h_{\alpha}(\eta)
\, . \label{muheta}
\end{align}
where $\alpha\in(0;2]$ and $\eta=\min\{\eta_{\cle},\eta_{\clt}\}$.
The result (\ref{muheta}) is an entropic uncertainty relation in
the model of detection inefficiencies. Entropies of actual
probability distributions take into account not only quantum
uncertainties. In the case $\alpha=1$, the inefficiency-free lower
bound is multiplied by $\eta$ and added by $2h_{1}(\eta)$.
Observations of similar kind were already reported in \cite{ramubs13}.

\section{Conclusions}\label{sec4}

It was emphasized by Pauli that a universal form of time operator
hardly exists. We have studied entropic uncertainty relations of the
``energy-time'' type on the basis of Pegg's concept of the
Hamiltonian complement. When ratios of energy values are rational
exactly or approximately, we can construct measurements with the
required properties. The Pegg concept allows us to treat the
energy-time uncertainty principle similarly to usual observables. It
also reflects features of time measurements, including possibly
arbitrary choice of reference moments. To express quantum
uncertainties, R\'{e}nyi and Tsallis entropies were utilized. The
derived uncertainty relations are immediately related to actual
measurement statistics. Since our relations characterize energy-time
uncertainties in more traditional style, they differ from recent
results reported in \cite{colesetal18}. On the other hand, obtained
entropic bounds of the Maassen--Uffink type turned out to be very
similar. In this regard, Pegg's concept of the Hamiltonian
complement leads to a supplementary treatment of the bounds
(\ref{renfr}) and (\ref{ctren}) within the preparation scenario.
Although our treatment is restricted to discrete levels of a certain
structure, it can sometimes be used in more general context. In many
cases, actual states of the principal system lie in a subspace
formed by some particular eigenstates of the Hamiltonian. That is,
one is subjected to a dynamical map that leaves this subspace
invariant. If the corresponding eigenvalues obey the condition to be
commensurable, then the derived uncertainty relations still hold.

In suitably extended space, the measurement of energy and its
complement can be treated as mutually unbiased. Hence, we derived
state-dependent entropic relations beyond the Maassen--Uffink
approach. Such bounds are expressed in terms of purity of the
measured quantum state. Another form of entropic uncertainty
relations follows from inequalities of the Landau--Pollak type.
Entropic uncertainty relations provide not only another way to
express some incompatibility of certain physical quantities. Such
relations may be of practical interest as imposing some restrictions
on probabilities of corresponding measurements. In this regard, the
question of detection inefficiencies was incorporated into a
consideration. Basic findings are similar to that was described
previously. Note also that state-dependent uncertainty relations of
various kind were formulated. In reality, inefficiency-free entropic
bounds will be multiplied by some factor depending on the efficiency
parameter. In addition, there are additive entropic terms related
purely to the employed detectors. It is known that entropic
uncertainty relations can be useful for information-processing
applications. Although the presented relations are restricted in
their scope, they are applicable to typical systems used for
information processing. Of course, many additional aspects of the
problem should be taken into account. As was mentioned above, the
role of dynamical effects in producing uncertainties is of certain
interest.

\end{document}